\documentstyle [epsfig,12pt]{article}

\newcommand{\gev}{\mbox{GeV}}

\newcommand{\kgam}{\mbox{$\kappa_{\gamma}$}}
\newcommand{\ggam}{\mbox{$g^{1}_{\gamma}$}} 
\newcommand{\deltak}{\mbox{$\Delta \kappa_{\gamma}$}}
\newcommand{\lk}{\mbox{$\lambda_{\gamma}$}}

\newcommand{\ptg}{\mbox{$p_{T\gamma}$}}
\newcommand{\ctg}{\mbox{$|\cos{\theta}_{\gamma}|$}}
\newcommand{\xe}{\mbox{$x_{E}$}}
\newcommand{\tg}{\mbox{$\theta_{\gamma}$}}
\newcommand{\eg}{\mbox{$E_{\gamma}$}}

\newcommand{\racs}{\mbox{$\sqrt s$}}

\newcommand{\egz}{\mbox{$E_{\gamma}^{\mathrm{Z}}$}}
\newcommand{\epm}{\mbox{$\mathrm{e}^{+} {e}^{-}$}}

\def\ap#1#2#3   {{\em Ann. Phys. (NY)} {\bf#1} (#2) #3.}
\def\apj#1#2#3  {{\em Astrophys. J.} {\bf#1} (#2) #3.}
\def\apjl#1#2#3 {{\em Astrophys. J. Lett.} {\bf#1} (#2) #3.}
\def\app#1#2#3  {{\em Acta. Phys. Pol.} {\bf#1} (#2) #3.}
\def\ar#1#2#3   {{\em Ann. Rev. Nucl. Part. Sci.} {\bf#1} (#2) #3.}
\def\eur#1#2#3  {{\em Eur. Phys. Journal} {\bf#1} (#2) #3.}
\def\cpc#1#2#3  {{\em Computer Phys. Comm.} {\bf#1} (#2) #3.}
\def\err#1#2#3  {{\it Erratum} {\bf#1} (#2) #3.}
\def\ib#1#2#3   {{\it ibid.} {\bf#1} (#2) #3.}
\def\jmp#1#2#3  {{\em J. Math. Phys.} {\bf#1} (#2) #3.}
\def\ijmp#1#2#3 {{\em Int. J. Mod. Phys.} {\bf#1} (#2) #3.}
\def\jetp#1#2#3 {{\em JETP Lett.} {\bf#1} (#2) #3.}
\def\jpg#1#2#3  {{\em J. Phys. G.} {\bf#1} (#2) #3.}
\def\mpl#1#2#3  {{\em Mod. Phys. Lett.} {\bf#1} (#2) #3.}
\def\nat#1#2#3  {{\em Nature (London)} {\bf#1} (#2) #3.}
\def\nc#1#2#3   {{\em Nuovo Cim.} {\bf#1} (#2) #3.}
\def\nim#1#2#3  {{\em Nucl. Instr. Meth.} {\bf#1} (#2) #3.}
\def\np#1#2#3   {{\em Nucl. Phys.} {\bf#1} (#2) #3.}
\def\pcps#1#2#3 {{\em Proc. Cam. Phil. Soc.} {\bf#1} (#2) #3.}
\def\pl#1#2#3   {{\em Phys. Lett.} {\bf#1} (#2) #3.}
\def\prep#1#2#3 {{\em Phys. Rep.} {\bf#1} (#2) #3.}
\def\prev#1#2#3 {{\em Phys. Rev.} {\bf#1} (#2) #3.}
\def\prl#1#2#3  {{\em Phys. Rev. Lett.} {\bf#1} (#2) #3.}
\def\prs#1#2#3  {{\em Proc. Roy. Soc.} {\bf#1} (#2) #3.}
\def\ptp#1#2#3  {{\em Prog. Th. Phys.} {\bf#1} (#2) #3.}
\def\ps#1#2#3   {{\em Physica Scripta} {\bf#1} (#2) #3.}
\def\rmp#1#2#3  {{\em Rev. Mod. Phys.} {\bf#1} (#2) #3.}
\def\rpp#1#2#3  {{\em Rep. Prog. Phys.} {\bf#1} (#2) #3.}
\def\sjnp#1#2#3 {{\em Sov. J. Nucl. Phys.} {\bf#1} (#2) #3.}
\def\spj#1#2#3  {{\em Sov. Phys. JEPT} {\bf#1} (#2) #3.}
\def\spu#1#2#3  {{\em Sov. Phys.-Usp.} {\bf#1} (#2) #3.}
\def\zp#1#2#3   {{\em Zeit. Phys.} {\bf#1} (#2) #3.}

\begin{document}
\begin{titlepage}
\begin{center}
{\large EUROPEAN LABORATORY FOR PARTICLES PHYSICS (CERN)}
\end{center}
\vspace{1.0 cm}
\vspace{1.0 cm}
\begin{flushright}
CERN-EP/98-178\\
16 November 1998 \\
\vspace{1.0 cm}

\end{flushright}
\vspace{1.5 cm}
\begin{center}

{\Large\bf Measurement of triple gauge
 WW$\gamma$  couplings at LEP2
using photonic events. 
}\\
\vspace{1.5cm}
{\bf  The ALEPH Collaboration  $^{*})$}\\

\end{center}
\vspace{3.0 cm}
 
\begin{abstract}
 
A study of  events with photons and missing energy has been performed with  the data sample obtained with the ALEPH detector
at centre-of-mass energies from 161 to 184 GeV, corresponding to a total integrated luminosity of about 80 $\mathrm{pb}^{-1}$.
The measured distributions are in agreement with Standard Model predictions, leading to constraints on 
 WW$\gamma$ gauge coupling parameters $\deltak$   and $\lk$. 
 The results from
the fit to the cross sections and to the  energy and angular distributions of the photons are:

\begin{eqnarray*}
       \deltak & = & \quad 0.05 ^{+1.15}_{-1.10} \mathrm{(stat)}  \pm 0.25         \mathrm{(syst)} \\
       \lk  & = & - \, 0.05  ^{+1.55}_{-1.45} \mathrm{(stat)}  \pm 0.30         \mathrm{(syst)}.
\end{eqnarray*}

\end{abstract}

\vspace*{1.5 cm}

\begin{center}
\it{(Submitted to Physics Letters B)}
\end{center}

\vspace*{1.5 cm}
\vfill
-------------------------------\\
$^*)$ {\small See next pages for the list of authors}

\pagestyle{empty}
\newpage
\small
%
%
%
%
\setlength{\parskip}{0.0cm}
\setlength{\textheight}{25.0cm}
\setlength{\topmargin}{-1.5cm}
\setlength{\textwidth}{16 cm}
\setlength{\oddsidemargin}{-0.0cm}
\setlength{\topsep}{1mm}
\pretolerance=10000
\centerline{\large\bf The ALEPH Collaboration}
\footnotesize
\vspace{0.5cm}
{\raggedbottom
\begin{sloppypar}
\samepage\noindent
R.~Barate,
D.~Buskulic,
D.~Decamp,
P.~Ghez,
C.~Goy,
S.~Jezequel,
J.-P.~Lees,
A.~Lucotte,
F.~Martin,
E.~Merle,
\mbox{M.-N.~Minard},
\mbox{J.-Y.~Nief},
P.~Perrodo,
B.~Pietrzyk
\nopagebreak
\begin{center}
\parbox{15.5cm}{\sl\samepage
Laboratoire de Physique des Particules (LAPP), IN$^{2}$P$^{3}$-CNRS,
F-74019 Annecy-le-Vieux Cedex, France}
\end{center}\end{sloppypar}
\vspace{2mm}
\begin{sloppypar}
\noindent
R.~Alemany,
M.P.~Casado,
M.~Chmeissani,
J.M.~Crespo,
M.~Delfino,
E.~Fernandez,
M.~Fernandez-Bosman,
Ll.~Garrido,$^{15}$
E.~Graug\`{e}s,
A.~Juste,
M.~Martinez,
G.~Merino,
R.~Miquel,
Ll.M.~Mir,
P.~Morawitz,
A.~Pacheco,
I.C.~Park,
A.~Pascual,
I.~Riu,
F.~Sanchez
\nopagebreak
\begin{center}
\parbox{15.5cm}{\sl\samepage
Institut de F\'{i}sica d'Altes Energies, Universitat Aut\`{o}noma
de Barcelona, 08193 Bellaterra (Barcelona), E-Spain$^{7}$}
\end{center}\end{sloppypar}
\vspace{2mm}
\begin{sloppypar}
\noindent
A.~Colaleo,
D.~Creanza,
M.~de~Palma,
G.~Gelao,
G.~Iaselli,
G.~Maggi,
M.~Maggi,
S.~Nuzzo,
A.~Ranieri,
G.~Raso,
F.~Ruggieri,
G.~Selvaggi,
L.~Silvestris,
P.~Tempesta,
A.~Tricomi,$^{3}$
G.~Zito
\nopagebreak
\begin{center}
\parbox{15.5cm}{\sl\samepage
Dipartimento di Fisica, INFN Sezione di Bari, I-70126 Bari, Italy}
\end{center}\end{sloppypar}
\vspace{2mm}
\begin{sloppypar}
\noindent
X.~Huang,
J.~Lin,
Q. Ouyang,
T.~Wang,
Y.~Xie,
R.~Xu,
S.~Xue,
J.~Zhang,
L.~Zhang,
W.~Zhao
\nopagebreak
\begin{center}
\parbox{15.5cm}{\sl\samepage
Institute of High-Energy Physics, Academia Sinica, Beijing, The People's
Republic of China$^{8}$}
\end{center}\end{sloppypar}
\vspace{2mm}
\begin{sloppypar}
\noindent
D.~Abbaneo,
U.~Becker,$^{22}$
G.~Boix,$^{2}$
M.~Cattaneo,
F.~Cerutti,
V.~Ciulli,
G.~Dissertori,
H.~Drevermann,
R.W.~Forty,
M.~Frank,
F.~Gianotti,
A.W.~Halley,
J.B.~Hansen,
J.~Harvey,
P.~Janot,
B.~Jost,
I.~Lehraus,
O.~Leroy,
C.~Loomis,
P.~Maley,
P.~Mato,
A.~Minten,
L.~Moneta,$^{20}$
A.~Moutoussi,
N.~Qi,
F.~Ranjard,
L.~Rolandi,
D.~Rousseau,
D.~Schlatter,
M.~Schmitt,$^{1}$
O.~Schneider,
W.~Tejessy,
F.~Teubert,
I.R.~Tomalin,
E.~Tournefier,
M.~Vreeswijk,
H.~Wachsmuth
\nopagebreak
\begin{center}
\parbox{15.5cm}{\sl\samepage
European Laboratory for Particle Physics (CERN), CH-1211 Geneva 23,
Switzerland}
\end{center}\end{sloppypar}
\vspace{2mm}
\begin{sloppypar}
\noindent
Z.~Ajaltouni,
F.~Badaud
G.~Chazelle,
O.~Deschamps,
S.~Dessagne,
A.~Falvard,
C.~Ferdi,
P.~Gay,
C.~Guicheney,
P.~Henrard,
J.~Jousset,
B.~Michel,
S.~Monteil,
\mbox{J-C.~Montret},
D.~Pallin,
P.~Perret,
F.~Podlyski
\nopagebreak
\begin{center}
\parbox{15.5cm}{\sl\samepage
Laboratoire de Physique Corpusculaire, Universit\'e Blaise Pascal,
IN$^{2}$P$^{3}$-CNRS, Clermont-Ferrand, F-63177 Aubi\`{e}re, France}
\end{center}\end{sloppypar}
\vspace{2mm}
\begin{sloppypar}
\noindent
J.D.~Hansen,
J.R.~Hansen,
P.H.~Hansen,
B.S.~Nilsson,
B.~Rensch,
A.~W\"a\"an\"anen
\begin{center}
\parbox{15.5cm}{\sl\samepage
Niels Bohr Institute, 2100 Copenhagen, DK-Denmark$^{9}$}
\end{center}\end{sloppypar}
\vspace{2mm}
\begin{sloppypar}
\noindent
G.~Daskalakis,
A.~Kyriakis,
C.~Markou,
E.~Simopoulou,
A.~Vayaki
\nopagebreak
\begin{center}
\parbox{15.5cm}{\sl\samepage
Nuclear Research Center Demokritos (NRCD), GR-15310 Attiki, Greece}
\end{center}\end{sloppypar}
\vspace{2mm}
\begin{sloppypar}
\noindent
A.~Blondel,
\mbox{J.-C.~Brient},
F.~Machefert,
A.~Roug\'{e},
M.~Rumpf,
R.~Tanaka,
A.~Valassi,$^{6}$
H.~Videau
\nopagebreak
\begin{center}
\parbox{15.5cm}{\sl\samepage
Laboratoire de Physique Nucl\'eaire et des Hautes Energies, Ecole
Polytechnique, IN$^{2}$P$^{3}$-CNRS, \mbox{F-91128} Palaiseau Cedex, France}
\end{center}\end{sloppypar}
\vspace{2mm}
\begin{sloppypar}
\noindent
E.~Focardi,
G.~Parrini,
K.~Zachariadou
\nopagebreak
\begin{center}
\parbox{15.5cm}{\sl\samepage
Dipartimento di Fisica, Universit\`a di Firenze, INFN Sezione di Firenze,
I-50125 Firenze, Italy}
\end{center}\end{sloppypar}
\vspace{2mm}
\begin{sloppypar}
\noindent
R.~Cavanaugh,
M.~Corden,
C.~Georgiopoulos,
T.~Huehn,
D.E.~Jaffe
\nopagebreak
\begin{center}
\parbox{15.5cm}{\sl\samepage
Supercomputer Computations Research Institute,
Florida State University,
Tallahassee, FL 32306-4052, USA $^{13,14}$}
\end{center}\end{sloppypar}
\vspace{2mm}
\begin{sloppypar}
\noindent
A.~Antonelli,
G.~Bencivenni,
G.~Bologna,$^{4}$
F.~Bossi,
P.~Campana,
G.~Capon,
V.~Chiarella,
P.~Laurelli,
G.~Mannocchi,$^{5}$
F.~Murtas,
G.P.~Murtas,
L.~Passalacqua,
M.~Pepe-Altarelli$^{12}$
\nopagebreak
\begin{center}
\parbox{15.5cm}{\sl\samepage
Laboratori Nazionali dell'INFN (LNF-INFN), I-00044 Frascati, Italy}
\end{center}\end{sloppypar}
\vspace{2mm}
\begin{sloppypar}
\noindent
M.~Chalmers,
L.~Curtis,
J.G.~Lynch,
P.~Negus,
V.~O'Shea,
B.~Raeven,
C.~Raine,
D.~Smith,
P.~Teixeira-Dias,
A.S.~Thompson,
E.~Thomson,
J.J.~Ward
\nopagebreak
\begin{center}
\parbox{15.5cm}{\sl\samepage
Department of Physics and Astronomy, University of Glasgow, Glasgow G12
8QQ,United Kingdom$^{10}$}
\end{center}\end{sloppypar}
\pagebreak
\begin{sloppypar}
\noindent
O.~Buchm\"uller,
S.~Dhamotharan,
C.~Geweniger,
P.~Hanke,
G.~Hansper,
V.~Hepp,
E.E.~Kluge,
A.~Putzer,
J.~Sommer,
K.~Tittel,
S.~Werner,$^{22}$
M.~Wunsch
\nopagebreak
\begin{center}
\parbox{15.5cm}{\sl\samepage
Institut f\"ur Hochenergiephysik, Universit\"at Heidelberg, D-69120
Heidelberg, Germany$^{16}$}
\end{center}\end{sloppypar}
\vspace{2mm}
\begin{sloppypar}
\noindent
R.~Beuselinck,
D.M.~Binnie,
W.~Cameron,
P.J.~Dornan,$^{12}$
M.~Girone,
S.~Goodsir,
N.~Marinelli,
E.B.~Martin,
J.~Nash,
J.K.~Sedgbeer,
P.~Spagnolo,
M.D.~Williams
\nopagebreak
\begin{center}
\parbox{15.5cm}{\sl\samepage
Department of Physics, Imperial College, London SW7 2BZ,
United Kingdom$^{10}$}
\end{center}\end{sloppypar}
\vspace{2mm}
\begin{sloppypar}
\noindent
V.M.~Ghete,
P.~Girtler,
E.~Kneringer,
D.~Kuhn,
G.~Rudolph
\nopagebreak
\begin{center}
\parbox{15.5cm}{\sl\samepage
Institut f\"ur Experimentalphysik, Universit\"at Innsbruck, A-6020
Innsbruck, Austria$^{18}$}
\end{center}\end{sloppypar}
\vspace{2mm}
\begin{sloppypar}
\noindent
A.P.~Betteridge,
C.K.~Bowdery,
P.G.~Buck,
P.~Colrain,
G.~Crawford,
G.~Ellis,
A.J.~Finch,
F.~Foster,
G.~Hughes,
R.W.L.~Jones,
A.N.~Robertson,
M.I.~Williams
\nopagebreak
\begin{center}
\parbox{15.5cm}{\sl\samepage
Department of Physics, University of Lancaster, Lancaster LA1 4YB,
United Kingdom$^{10}$}
\end{center}\end{sloppypar}
\vspace{2mm}
\begin{sloppypar}
\noindent
P.~van~Gemmeren,
I.~Giehl,
C.~Hoffmann,
K.~Jakobs,
K.~Kleinknecht,
M.~Kr\"ocker,
H.-A.~N\"urnberger,
G.~Quast,
B.~Renk,
E.~Rohne,
H.-G.~Sander,
S.~Schmeling,
C.~Zeitnitz,
T.~Ziegler
\nopagebreak
\begin{center}
\parbox{15.5cm}{\sl\samepage
Institut f\"ur Physik, Universit\"at Mainz, D-55099 Mainz, Germany$^{16}$}
\end{center}\end{sloppypar}
\vspace{2mm}
\begin{sloppypar}
\noindent
J.J.~Aubert,
C.~Benchouk,
A.~Bonissent,
J.~Carr,$^{12}$
P.~Coyle,
A.~Ealet,
D.~Fouchez,
F.~Motsch,
P.~Payre,
M.~Talby,
M.~Thulasidas,
A.~Tilquin
\nopagebreak
\begin{center}
\parbox{15.5cm}{\sl\samepage
Centre de Physique des Particules, Facult\'e des Sciences de Luminy,
IN$^{2}$P$^{3}$-CNRS, F-13288 Marseille, France}
\end{center}\end{sloppypar}
\vspace{2mm}
\begin{sloppypar}
\noindent
M.~Aleppo,
M.~Antonelli,
F.~Ragusa
\nopagebreak
\begin{center}
\parbox{15.5cm}{\sl\samepage
Dipartimento di Fisica, Universit\`a di Milano e INFN Sezione di
Milano, I-20133 Milano, Italy.}
\end{center}\end{sloppypar}
\vspace{2mm}
\begin{sloppypar}
\noindent
R.~Berlich,
V.~B\"uscher,
H.~Dietl,
G.~Ganis,
K.~H\"uttmann,
G.~L\"utjens,
C.~Mannert,
W.~M\"anner,
\mbox{H.-G.~Moser},
S.~Schael,
R.~Settles,
H.~Seywerd,
H.~Stenzel,
W.~Wiedenmann,
G.~Wolf
\nopagebreak
\begin{center}
\parbox{15.5cm}{\sl\samepage
Max-Planck-Institut f\"ur Physik, Werner-Heisenberg-Institut,
D-80805 M\"unchen, Germany\footnotemark[16]}
\end{center}\end{sloppypar}
\vspace{2mm}
\begin{sloppypar}
\noindent
P.~Azzurri,
J.~Boucrot,
O.~Callot,
S.~Chen,
M.~Davier,
L.~Duflot,
\mbox{J.-F.~Grivaz},
Ph.~Heusse,
A.~Jacholkowska,
M.~Kado,
J.~Lefran\c{c}ois,
L.~Serin,
\mbox{J.-J.~Veillet},
I.~Videau,$^{12}$
J.-B.~de~Vivie~de~R\'egie,
D.~Zerwas
\nopagebreak
\begin{center}
\parbox{15.5cm}{\sl\samepage
Laboratoire de l'Acc\'el\'erateur Lin\'eaire, Universit\'e de Paris-Sud,
IN$^{2}$P$^{3}$-CNRS, F-91898 Orsay Cedex, France}
\end{center}\end{sloppypar}
\vspace{2mm}
\begin{sloppypar}
\noindent
\samepage
G.~Bagliesi,$^{12}$
S.~Bettarini,
T.~Boccali,
C.~Bozzi,
G.~Calderini,
R.~Dell'Orso,
I.~Ferrante,
A.~Giassi,
A.~Gregorio,
F.~Ligabue,
A.~Lusiani,
P.S.~Marrocchesi,
A.~Messineo,
F.~Palla,
G.~Rizzo,
G.~Sanguinetti,
A.~Sciab\`a,
G.~Sguazzoni,
R.~Tenchini,
C.~Vannini,
A.~Venturi,
P.G.~Verdini
\samepage
\begin{center}
\parbox{15.5cm}{\sl\samepage
Dipartimento di Fisica dell'Universit\`a, INFN Sezione di Pisa,
e Scuola Normale Superiore, I-56010 Pisa, Italy}
\end{center}\end{sloppypar}
\vspace{2mm}
\begin{sloppypar}
\noindent
G.A.~Blair,
J.T.~Chambers,
J.~Coles,
G.~Cowan,
M.G.~Green,
T.~Medcalf,
J.A.~Strong,
J.H.~von~Wimmersperg-Toeller
\nopagebreak
\begin{center}
\parbox{15.5cm}{\sl\samepage
Department of Physics, Royal Holloway \& Bedford New College,
University of London, Surrey TW20 OEX, United Kingdom$^{10}$}
\end{center}\end{sloppypar}
\vspace{2mm}
\begin{sloppypar}
\noindent
D.R.~Botterill,
R.W.~Clifft,
T.R.~Edgecock,
P.R.~Norton,
J.C.~Thompson,
A.E.~Wright
\nopagebreak
\begin{center}
\parbox{15.5cm}{\sl\samepage
Particle Physics Dept., Rutherford Appleton Laboratory,
Chilton, Didcot, Oxon OX11 OQX, United Kingdom$^{10}$}
\end{center}\end{sloppypar}
\vspace{2mm}
\begin{sloppypar}
\noindent
\mbox{B.~Bloch-Devaux},
P.~Colas,
B.~Fabbro,
G.~Fa\"\i f,
E.~Lan\c{c}on,$^{12}$
\mbox{M.-C.~Lemaire},
E.~Locci,
P.~Perez,
H.~Przysiezniak,
J.~Rander,
\mbox{J.-F.~Renardy},
A.~Rosowsky,
A.~Trabelsi,$^{23}$
B.~Tuchming,
B.~Vallage
\nopagebreak
\begin{center}
\parbox{15.5cm}{\sl\samepage
CEA, DAPNIA/Service de Physique des Particules,
CE-Saclay, F-91191 Gif-sur-Yvette Cedex, France$^{17}$}
\end{center}\end{sloppypar}
\vspace{2mm}
\begin{sloppypar}
\noindent
S.N.~Black,
J.H.~Dann,
H.Y.~Kim,
N.~Konstantinidis,
A.M.~Litke,
M.A. McNeil,
G.~Taylor
\nopagebreak
\begin{center}
\parbox{15.5cm}{\sl\samepage
Institute for Particle Physics, University of California at
Santa Cruz, Santa Cruz, CA 95064, USA$^{19}$}
\end{center}\end{sloppypar}
\pagebreak
\vspace{2mm}
\begin{sloppypar}
\noindent
C.N.~Booth,
S.~Cartwright,
F.~Combley,
M.S.~Kelly,
M.~Lehto,
L.F.~Thompson
\nopagebreak
\begin{center}
\parbox{15.5cm}{\sl\samepage
Department of Physics, University of Sheffield, Sheffield S3 7RH,
United Kingdom$^{10}$}
\end{center}\end{sloppypar}
\vspace{2mm}
\begin{sloppypar}
\noindent
K.~Affholderbach,
A.~B\"ohrer,
S.~Brandt,
J.~Foss,
C.~Grupen,
G.~Prange,
L.~Smolik,
F.~Stephan
\nopagebreak
\begin{center}
\parbox{15.5cm}{\sl\samepage
Fachbereich Physik, Universit\"at Siegen, D-57068 Siegen, Germany$^{16}$}
\end{center}\end{sloppypar}
\vspace{2mm}
\begin{sloppypar}
\noindent
G.~Giannini,
B.~Gobbo
\nopagebreak
\begin{center}
\parbox{15.5cm}{\sl\samepage
Dipartimento di Fisica, Universit\`a di Trieste e INFN Sezione di Trieste,
I-34127 Trieste, Italy}
\end{center}\end{sloppypar}
\vspace{2mm}
\begin{sloppypar}
\noindent
J.~Putz,
J.~Rothberg,
S.~Wasserbaech,
R.W.~Williams
\nopagebreak
\begin{center}
\parbox{15.5cm}{\sl\samepage
Experimental Elementary Particle Physics, University of Washington, WA 98195
Seattle, U.S.A.}
\end{center}\end{sloppypar}
\vspace{2mm}
\begin{sloppypar}
\noindent
S.R.~Armstrong,
E.~Charles,
P.~Elmer,
D.P.S.~Ferguson,
Y.~Gao,
S.~Gonz\'{a}lez,
T.C.~Greening,
O.J.~Hayes,
H.~Hu,
S.~Jin,
P.A.~McNamara III,
J.M.~Nachtman,$^{21}$
J.~Nielsen,
W.~Orejudos,
Y.B.~Pan,
Y.~Saadi,
I.J.~Scott,
J.~Walsh,
Sau~Lan~Wu,
X.~Wu,
G.~Zobernig
\nopagebreak
\begin{center}
\parbox{15.5cm}{\sl\samepage
Department of Physics, University of Wisconsin, Madison, WI 53706,
USA$^{11}$}
\end{center}\end{sloppypar}
}
\footnotetext[1]{Now at Harvard University, Cambridge, MA 02138, U.S.A.}
\footnotetext[2]{Supported by the Commission of the European Communities,
contract ERBFMBICT982894.}
\footnotetext[3]{Also at Dipartimento di Fisica, INFN Sezione di Catania,
Catania, Italy.}
\footnotetext[4]{Also Istituto di Fisica Generale, Universit\`{a} di
Torino, Torino, Italy.}
\footnotetext[5]{Also Istituto di Cosmo-Geofisica del C.N.R., Torino,
Italy.}
\footnotetext[6]{Now at LAL, Orsay, France.}
\footnotetext[7]{Supported by CICYT, Spain.}
\footnotetext[8]{Supported by the National Science Foundation of China.}
\footnotetext[9]{Supported by the Danish Natural Science Research Council.}
\footnotetext[10]{Supported by the UK Particle Physics and Astronomy Research
Council.}
\footnotetext[11]{Supported by the US Department of Energy, grant
DE-FG0295-ER40896.}
\footnotetext[12]{Also at CERN, 1211 Geneva 23, Switzerland.}
\footnotetext[13]{Supported by the US Department of Energy, contract
DE-FG05-92ER40742.}
\footnotetext[14]{Supported by the US Department of Energy, contract
DE-FC05-85ER250000.}
\footnotetext[15]{Permanent address: Universitat de Barcelona, 08208 Barcelona,
Spain.}
\footnotetext[16]{Supported by the Bundesministerium f\"ur Bildung,
Wissenschaft, Forschung und Technologie, Germany.}
\footnotetext[17]{Supported by the Direction des Sciences de la
Mati\`ere, C.E.A.}
\footnotetext[18]{Supported by Fonds zur F\"orderung der wissenschaftlichen
Forschung, Austria.}
\footnotetext[19]{Supported by the US Department of Energy,
grant DE-FG03-92ER40689.}
\footnotetext[20]{Now at University of Geneva, 1211 Geneva 4, Switzerland.}
\footnotetext[21]{Now at University of California at Los Angeles (UCLA),
Los Angeles, CA 90024, U.S.A.}
\footnotetext[22]{Now at SAP AG, D-69185 Walldorf, Germany}
\footnotetext[23]{Now at D\'epartement de Physique, Facult\'e des Sciences de Tunis, 1060 Le Belv\'ed\`ere, Tunisia.}
%
%
\normalsize
\newpage
\pagestyle{plain}
\setcounter{page}{1}

\end{titlepage}

\pagenumbering{arabic}
\setcounter{section}{0}
\setcounter{equation}{0}
\raggedbottom

\setlength{\textheight}{24.0cm}
\setlength{\topmargin}{-0.5cm}
\setlength{\textwidth}{15.0cm}
\setlength{\oddsidemargin}{+0.8cm}
\setlength{\topsep}{1mm}
\parskip= 3pt plus 1pt
\clearpage
\setcounter{totalnumber}{5}

\renewcommand{\textfraction}{0.1}
\renewcommand{\floatpagefraction}{0.8}
\renewcommand{\topfraction}{0.9}
\renewcommand{\bottomfraction}{0.9}

\vskip 1.0cm 
\section{Introduction}
 
In $\epm$ collisions at LEP 2 energies, the trilinear WW$\gamma$ and WWZ  couplings can be
probed with direct W-pair
 ($\epm \rightarrow  \mathrm{W^{+} W^{-}} $), single W 
 ($\epm     \rightarrow \mathrm{W e} \nu $)
 production or with photon production 
($\epm    \rightarrow \nu \overline{\nu} \gamma (\gamma) $ ) [1,2].
In the WW channel a minimal set of five independent
parameters is necessary to describe the Z and $\gamma$ couplings to the W, assuming C and CP conservation.
Usually a
model-dependence is introduced to reduce this set to at most three parameters
(e.g. the model with the parameters
 $\alpha_{W} , \alpha_{W\phi} , \alpha_{B\phi}$ \cite{ref8}).
Although the photonic channel is less sensitive to the couplings than the
 W pair and single W channels \cite{refWW}, 
it can resolve sign ambiguities and is therefore
complementary. Constraints on the WW$\gamma$ vertex have also been obtained at the Tevatron \cite{D0} within 
              a slightly different theoretical framework.

 The purpose of this letter is to set constraints on the WW$\gamma$  trilinear
couplings with a study of  photonic events, using data collected
by ALEPH at centre-of-mass energies ranging from 161 to 184 GeV and corresponding 
to a total integrated luminosity of about 80 $\mathrm{pb}^{-1}$.

In the Standard Model, three processes contribute at tree level to the
$\nu \overline{\nu} \gamma $ final state corresponding to the five diagrams  
 shown in Figure 1.
 
\begin{figure}[htbp]
\begin{picture}(200,180)
\put(-120,-90){\epsfxsize200mm\epsfbox{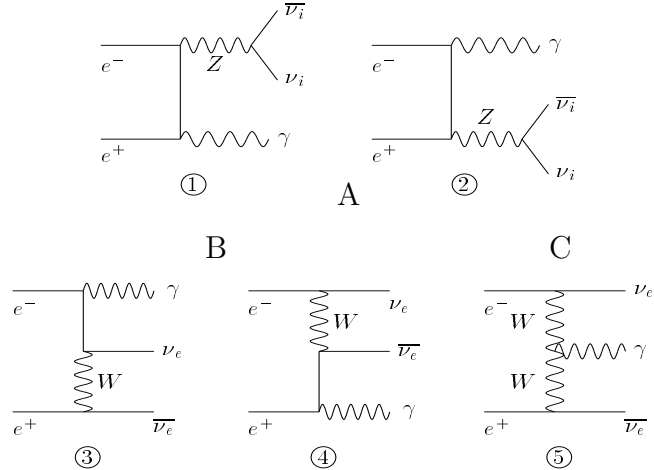}}
\put(210,100){A}
\put(160,80){B}
\put(290,80){C}
\end{picture}
 \caption{{Feynman diagrams for
          $\epm    \rightarrow \bar \nu \nu \gamma$.
          Only the process C is sensitive to the 
          WW$\gamma$ couplings.}}
 \label{f:diagrams}
\end{figure}

The WW$\gamma$ vertex is present only in the last diagram, which contributes
about 0.3\% to the total Standard Model  $\epm    \rightarrow \bar \nu \nu \gamma$ cross section, but which also
leads to characteristic energy and angular distributions of the final
state photons. A measurement of the total cross section, supplemented
by a fit to  these distributions, is therefore sensitive to the presence
of an anomalous WW$\gamma$ coupling.

This vertex can be described \cite{ref8} by three C and P conserving
parameters, $\ggam$, $\kgam$ and $\lk$, related to the following W boson
properties:

\begin{eqnarray*}
     \mathrm{charge}                      &       Q_{w}  =    &  e \ggam    \\
     \mathrm{magnetic \; dipole \; moment}        & \mu_{w}  =  &  \frac{e}{2 m_{w}} (\ggam + \kgam + \lk)   \\
     \mathrm{electric \; quadrupole \; moment}   &  q_{w} =   & -  \frac{e}{m_{w}^{2}}   ( \kgam - \lk).   
\end{eqnarray*}

In the Standard Model, these three parameters are equal to 1, 1 and 0,
res\-pec\-tively, and
their deviations from these values are parameterized as
``anomalous couplings'' $\Delta \ggam$, $\deltak$ and $\lk$. Here,
the electric charge of the W boson is assumed to be equal to that
of the electron, thus fixing $\ggam = 1$, while no further
assumptions are made on $\kgam$ and $\lk$. The matrix element for the
  $\epm    \rightarrow\nu \overline{\nu} \gamma (\gamma) $ final state
is a linear function of $\deltak$ and $\lk$. 
Its implementation in the KORALZ Monte Carlo program \cite{was}, including initial state radiation of additional photons,
is used throughout this analysis.

This letter is organized as follows. In Section 2, the aspects of
the ALEPH detector relevant to this analysis are described. The event
selection is presented in Section 3, and the fit of the data to the
presence of anomalous WW$\gamma$ couplings is discussed in Section 4. Section
5 gives the fitted values and the resulting constraints on $\deltak$
and $\lk$; 
 the systematic uncertainties are discussed in Section 6.

\section{The ALEPH detector}
The ALEPH detector and its performance are described in detail in
\cite{aleph,perf}. Here only a brief description of the properties relevant to the present analysis is given. 
 
The central part is dedicated to the detection 
of charged particles. From the interaction point
outwards, the trajectory of a charged particle is measured by a two-layer silicon strip vertex detector,
a cylindrical drift chamber and a large time projection chamber (TPC). The three
tracking detectors are immersed in a $1.5$ T axial field provided by a superconducting
solenoidal coil.

Photons are identified in the electromagnetic calorimeter (ECAL), situated bet\-ween the TPC and the coil. It is a 
lead--proportional--wire sampling calorimeter segmented in $0.9^{\circ} \times  0.9^{\circ}$ towers  read out 
in three sections in depth. It has  
 a total thickness of 22 radiation lengths
 and yields an energy  resolution of $\delta E/E = 0.18/\sqrt{E} +0.009$ ($E$ in
\gev).  Two independent readouts of the energy are implemented respectively on the cathode pads and on the anode wires of the ECAL.
At low polar angles, the ECAL is supplemented by two calorimeters, LCAL and SiCAL, principally used to measure the integrated
luminosity collected by the experiment, but used also here for vetoeing purposes. 

The iron return yoke  is equipped with 23 layers of streamer tubes and forms the
hadron calorimeter (HCAL), seven interaction lengths thick; it provides a
relative energy resolution of charged and neutral hadrons of $0.85/\sqrt{E}$.    
 Muons are identified using hits  in the HCAL
and the muon chambers; the latter  are composed of two layers of streamer tubes outside the HCAL.
  
 The information from the tracking detectors and the calorimeters are combined in an
energy flow algorithm \cite{perf}. For each event, the algorithm provides a
set of charged and neutral reconstructed particles, called energy flow objects, used in the analysis.

\section{Event samples and selection}

The data were collected with the ALEPH detector at LEP  at several centre-of-mass
 energies between 161 and 172 GeV in 1996, and between 181 and 184 GeV in 1997.
The corresponding integrated luminosities are given in 
Table \ref {t:dats}.

\begin{table}[htbp]
\caption{Data  samples.}
\begin{center}
\begin{tabular}{|l|c|c|} \hline
Energy                 &   Luminosity           &       N events                             \\
(GeV)                  &   ($\mathrm{pb^{-1}}$) & Data \quad  Expected                         \\ \hline \hline
 161                   &   11.0                 &  32  \qquad   31.8                           \\
 172                   &   10.7                 &  27  \qquad   32.2                           \\
 183                   &   58.1                 & 148  \qquad   145.8                          \\
 Total                 &   79.8                 & 207  \qquad   209.8                          \\ \hline
\end{tabular}
\label{t:dats}
\end{center}
\end{table}

\subsection{Selections and cuts} 

 Photon candidates are defined as described in \cite{perf}.
Only  events with no reconstructed charged particle tracks and at least one photon with an energy
  $ \eg   >   0.1 \racs$  
 are considered; the trigger efficiency for such events is almost 100\%. 
At most one hit is  accepted in the muon chambers, to eliminate beam-related and cosmic ray muons.
The loss of signal events with noisy muon chambers was estimated from events triggered at random
beam crossings to be $3 \%$.
 The  timing of the energy deposition in the ECAL  is checked  to be consistent with the beam crossing time.      
 
All events with at least 0.5 GeV
detected below  $14^{\circ}$ from the beam axis
are rejected, in order to remove radiative Bhabha events.
The efficiency correction factor associated with this cut was  estimated from events triggered at random beam crossings 
to be $3.5\%$.

The consistency between the energy measured from the ECAL pads and from the ECAL wires is checked.
In case of leakage out of the ECAL, a localized energy deposit in the HCAL, $E_{had}$, associated to an ECAL cluster is
added to $\eg$, after correcting   for the $\mathrm{e/\pi}$ ratio; only events with  $E_{had}/\eg < 10 \%$ are kept.
 To reduce the remaining background,  all but 2.5 GeV of the total energy             
  is required to come from photon candidates.

 At least one photon candidate is required to fulfil the  conditions 
 $\tg > 20^{\circ}$  and
 $ \ptg / E_{beam} > 0.1$.
  For multiphoton candidates,
the additional photons are considered only if their energy exceeds 
$  0.05 \racs$.  
The overall missing transverse momentum 
is required to be greater than  12 GeV/{\it c}. The last cut removes the 
 remaining  Bhabha events with radiation at large angle.
 
 Table  \ref {t:dats} shows the data samples used
 in this analysis.
 The numbers of selected events agree  with the numbers expected from the SM cross sections determined with 
 the KORALZ Monte Carlo.  The cross section measured from the data 
at 183 GeV, with the present analysis and within the global kinematic cuts, 
is $3.45 \pm 0.30$ pb, to be compared to the SM
prediction of $3.40 \pm 0.02$ pb.

\subsection{Monte Carlo simulation with KORALZ}

The simulation uses a modified version of the
 KORALZ program, which includes the
SM expectation (with electroweak corrections) as well as QED radiative
corrections, and the contributions of anomalous coupling amplitudes with exact matrix element calculations
 \cite{kalinowski}. The overall higher-order QED correction factor is around 1.4, but depends on the centre-of-mass energy.
More than ten thousand simulated events are used for each energy.
  
To obtain a description of the anomalous couplings in the simulation,  each event is assigned a
weight, which is a function of $\deltak$ and $\lk$.
This method  provides the smallest uncertainty, as the statistical error corresponds only to the differences between
the distributions produced from the Standard Model and those  from anomalous matrix elements.

\begin{figure}[b]
\begin{center}
\mbox{\epsfig{file=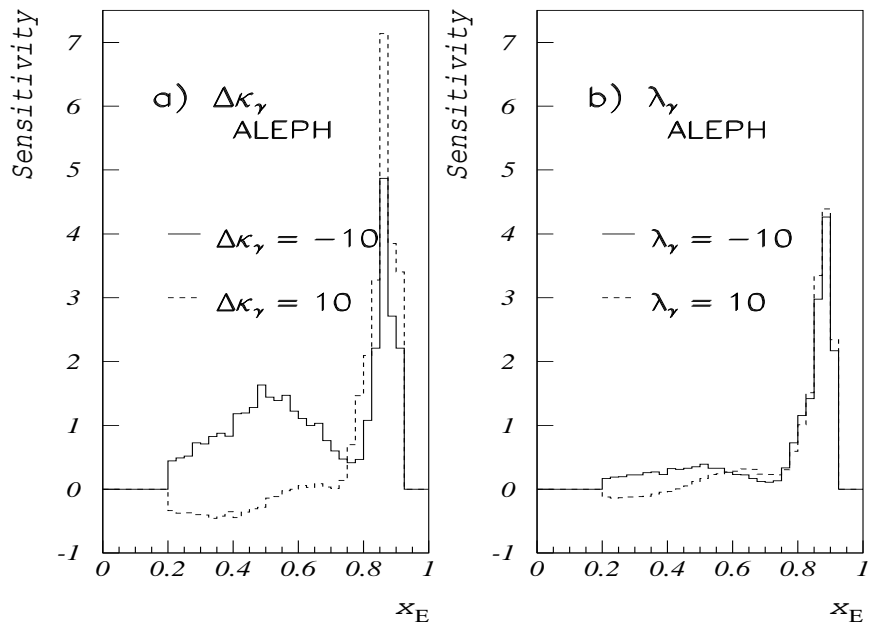,
width=13cm,height=10cm}}
\end{center}
\caption{Statistical sensitivity of this analysis as a function of the scaled energy $\xe$ at 183 GeV centre-of-mass energy:
a) for $\deltak$,  b) for $\lk$.
The sensitivity is defined as the anomalous contribution divided by the statistical error on the SM expectation.
The solid and dashed histograms correspond to  parameter values of $-10$ and  +10, respectively.
The single photon radiative return to the Z corresponds to $ \xe = 0.75$.
}
\label{f:sensix}
\end{figure}

As the matrix element is linear in $\deltak$ and $\lk$
 the cross section and the differential distributions are bilinear forms of $\deltak$ and $\lk$.
 For each event it is thus sufficient to store weights for only six configurations in the ($\deltak , \lk$) plane,
 in order to compute any cross section or kinematic variable as a function of $\deltak$ and $\lk$.
 The leading order amplitudes including these anomalous couplings are folded with higher order QED effects,
 following the procedure discussed in \cite{reffut}.

The Standard Model predicts that the cross section for  the radiative return to the Z resonance
 decreases when the centre-of-mass energy increases, while the opposite is true for the
W exchange.
In case of anomalous contributions, the  sensitivity of the cross section
 increases almost quadratically above 161 GeV.

The kinematic cuts have been chosen to optimize  
  the sensitivity to the anomalous couplings $\deltak$ and $\lk$.
Figure \ref {f:sensix} gives  the statistical sensitivity, defined as the anomalous contribution divided by the statistical error
on the SM expectation,
as a function of the scaled energy variable $\xe = \eg / E_{beam}$.
The minimum of the sensitivity occurs around the position of the Z return peak ($\xe = 0.75$).
  An important observation
 is that for $\deltak >0$ the differential cross section decreases almost linearly
  as a function of $\deltak$ for events with $\xe < 0.75$,  and increases quadratically above.

\begin{figure}[t]
\begin{center}
\mbox{\epsfig{file=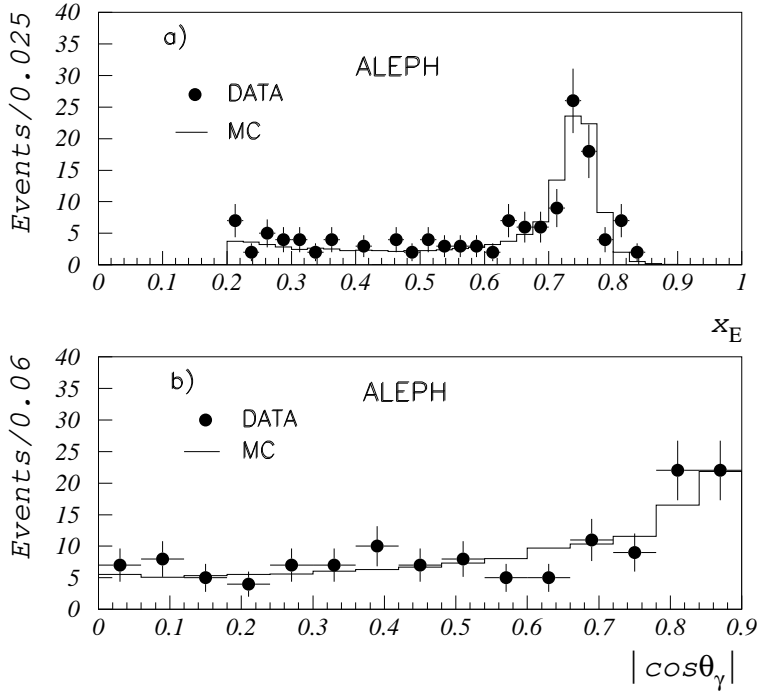,
width=11cm,height=11cm}}
\end{center}
\caption{Inclusive  distribution of a) the scaled energy $\xe$ and b) the absolute value of the cosine of the polar angle
of the photons, after all selections, for data and Monte Carlo at 183 GeV.}
\label{f:xescaled}
\end{figure}

\section{Likelihood fit}

In addition to the observed number of events, two kinematic variables of the photon are used in the fit:
the photon polar angle $\theta_{\gamma}$ and
 the scaled energy $\xe$.
 Figure \ref {f:xescaled}  shows the distribution of $\xe$
 and $\ctg$ for data, compared to the Standard Model predictions for $\racs = 183$ GeV.

 Two kinematic regions have been chosen for the fit, exclu\-ding the region of the Z peak return 
where the sensitivity to the anomalous couplings is minimal.
Only photons with $\ctg < 0.9$ are used for the fit to the shape of the distributions.

 Defining $ \egz = (s-m^2_{\mathrm{Z}})/2\racs$,
 the two kinematic regions are the following:

\begin{itemize}
\item  Region 1, low energy photons with $\eg < \egz - 3\Gamma_{\mathrm{Z}} $

 The contribution from higher order radiative corrections
 is described  by an almost  constant term obtained from the Monte Carlo simulation.
The scaled  variable $\xe$ is
found to be as discriminant as the angular variable in the fit. Both are used for the $\deltak$ fit, whereas $\lk$ is
determined only from the total cross section. The sensitivity to the $\lk$ parameter in this kinematic region is very low.

\item  Region 2, high energy photons with $\eg  > \egz + 0.5 \quad \mathrm{GeV}$

 In this region, the  higher order radiative corrections
 decrease the number of expected events by 30\%.
The scaled energy $\xe$
is more discriminant than the angular variable,
both variables being used in the fit of $\deltak$ and $\lk$. It can be observed (Figure  \ref {f:sensix})  that
the sensitivity to  $\lk$ with $\xe$  is similar to that of $\deltak$.
\end{itemize}

  Limits for anomalous coupling parameters have been derived from the generalized likelihood expression:

$
 \mathrm{log}L = \mathrm{log}\frac{\mathrm{(N_{th}^{(1)})^{N_{obs}^{(1)}}} \,
                  e^{-N_{th}^{(1)}}}{\mathrm{N_{obs}^{(1)}} !}
        + \mathrm{log}\frac{\mathrm{(N_{th}^{(2)})^{N_{obs}^{(2)}}} \,
            e^{-N_{th}^{(2)}}}{\mathrm{N_{obs}^{(2)}} !}
        +\sum logP_{i}^{(1)}
        +\sum logP_{i}^{(2)},
$

where $\mathrm{P_{i}^{(1)}}$, $\mathrm{P_{i}^{(2)}}$ are
 the probability density functions of observing  event i with a given value of 
$ \xe$ and $ \tg$ in region 1 and 2 respectively,
and $\mathrm{N_{th}^{(1)}}$  and $\mathrm{N_{th}^{(2)}}$
are the expected number of events in each region, including background.
 This likelihood formula contains two parts: the first one concerning the number of observed events, the second one
 being related to differential distributions for each kinematic region.
The number of events used in the fit and those expected from the SM are given in Table 2.


\begin{table}[b]
\caption{Number of events (N Events) entering the fit in the two kinematic regions.
 The number of expected events is estimated from the KORALZ
cross sections, corrected for acceptance.}

\begin{center}
\begin{tabular}{|l|c|c|} \hline
   Kinematic                  & N Events,             & N Events,                                          \\
     region                   & Cross section fit    & ($\mathrm{x_{E}},\theta$) fit                     \\
                              & Data \quad   Expected & Data \quad   Expected                              \\ \hline \hline
                              &                       &                                                    \\
 Region 1                     &   93  \qquad  101.0    &    60    \qquad    67.4                           \\
 Region 2                     &    30 \qquad  32.8     &    23    \qquad    25.6                           \\  \hline
\end{tabular}
\label{t:kins}
\end{center}
\end{table}

The acceptance convoluted with the experimental resolution 
leads to correction factors to the cross sections of 1.10 for the
first  kinematic region and 0.7 for the second; these correction factors are constant (within $\pm 2 \%$)
in each region as $\deltak$ or $\lk$ vary.

\begin{figure}[t]
\begin{center}
\mbox{\epsfig{file=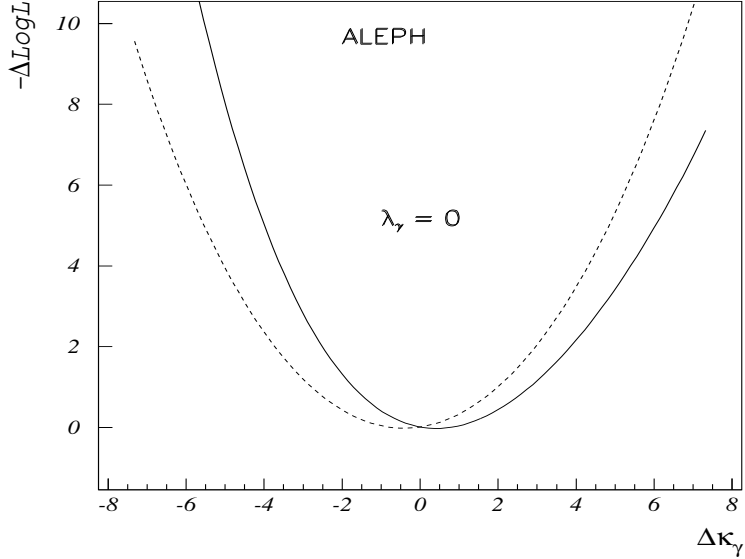,
width=11cm,height=9cm}}
\end{center}

\caption{Likelihood curves for  the fit of $\deltak$ at $\lk = 0$
for the contribution of the cross section term (solid curve) and the shape term in $\xe$ and $\theta$ (dashed curve).}

\label{f:likeli}
\end{figure}

\begin{figure}[t]
\begin{center}
\mbox{\epsfig{file=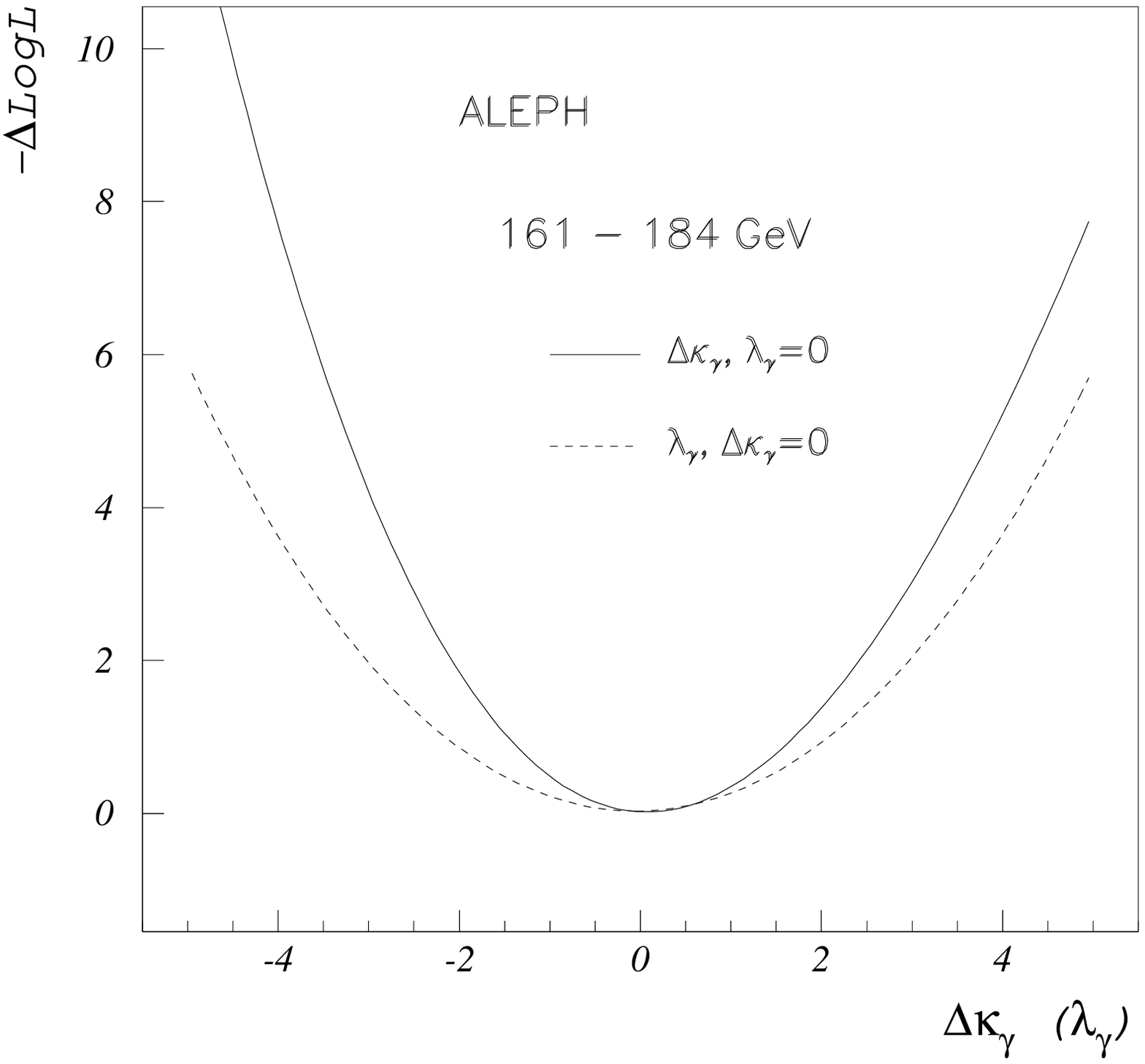,
width=11cm,height=9cm}}
\end{center}

\caption{Likelihood curves for
 the fit of $\lk$ at $\deltak = 0$ (solid curve) and $\deltak$ at $\lk = 0$
(dashed curve) for the sum of the cross section and distribution shape terms.}
\label{f:likeli2}
\end{figure}
 
    The studies made with the KORALZ Monte Carlo show that
   the cross sections  and distribution shapes vary
differently in the two kinematic regions.
For low energy photons the anomalous effects result from the interference term between the SM amplitude and the anomalous
amplitude; the resulting variation is monotonic  and linear for  $\deltak(\lk) > 0$
and  $\deltak(\lk) < 0$ and  only one solution is expected for the
 $\deltak$ and $\lk$ fit. For the high energy photons, the variations are quadratic (due to a quadratic contribution of the
 anomalous amplitude) and
one or two solutions are expected; the case of one solution corresponds to
  $\deltak = 0$ or $\lk = 0.$
This behaviour, important in the error determination,  
is discussed later when the error calibration procedure is presented.

\section{Results}
 
  The likelihood functions are       calculated globally for the cross section and on an event-by-event basis
 for the energy and angular distributions.
Figure \ref {f:likeli}  displays the variations of the log-likelihood ($-\Delta \mathrm{log} L$) 
 corresponding to the fit of $\deltak$
at $\lk = 0$, for the cross section and  the distribution contributions.
At present energies, the  contributions of the cross section and of the shape variation terms 
are equally important for the fit of $\deltak$.
The result for $\lk$ is dominated by the sensitivity to the shape in Region 2.
 
   Figure \ref {f:likeli2}  shows the $(-\Delta \mathrm{log}L)$ functions for  $\deltak$ fitted at
 $\lk = 0$, and for $\lk$ fitted at $\deltak = 0$ when the two contributions are merged.  The results are:
\begin{eqnarray*}
       \deltak & = & \quad 0.05    ^{+1.15}_{-1.10} \mathrm{(stat)}  \quad  \mathrm{assuming}  \quad \lk = 0  \\ 
        \lk  & = & - \, 0.05  ^{+1.55}_{-1.45} \mathrm{(stat)}  \quad  \mathrm{assuming}  \quad   \deltak = 0
\end{eqnarray*}
where the errors correspond to an increase of $-$logL by  0.5. 
   The lower precision for $\lk$ is expected since 
the exchanged W's are at a rather low momentum scale
 and  the  $\lk$ term in the Lagrangian
contains high powers of the W momentum.

The 95\% C.L. limits derived from the one parameter fits are :

\begin{eqnarray*}
       -2.1 < \deltak < 2.2   \quad  \mathrm{assuming}  \quad \lk = 0\\
       -3.0 <  \lk <  3.1      \quad  \mathrm{assuming}  \quad \deltak=0.
\end{eqnarray*} 

 The validity of these 95\% C.L. limits have been checked using 100 Monte Carlo samples 
corresponding to the data luminosity, the analysis procedure
described for the data being applied to each Monte Carlo sample. This study indicates that
these errors are  consistent with the frequentist interpretation, within 10\% of their values, and do not benefit 
from favourable statistical fluctuations.

\begin{figure}[t]
\begin{center}
\mbox{\epsfig{file=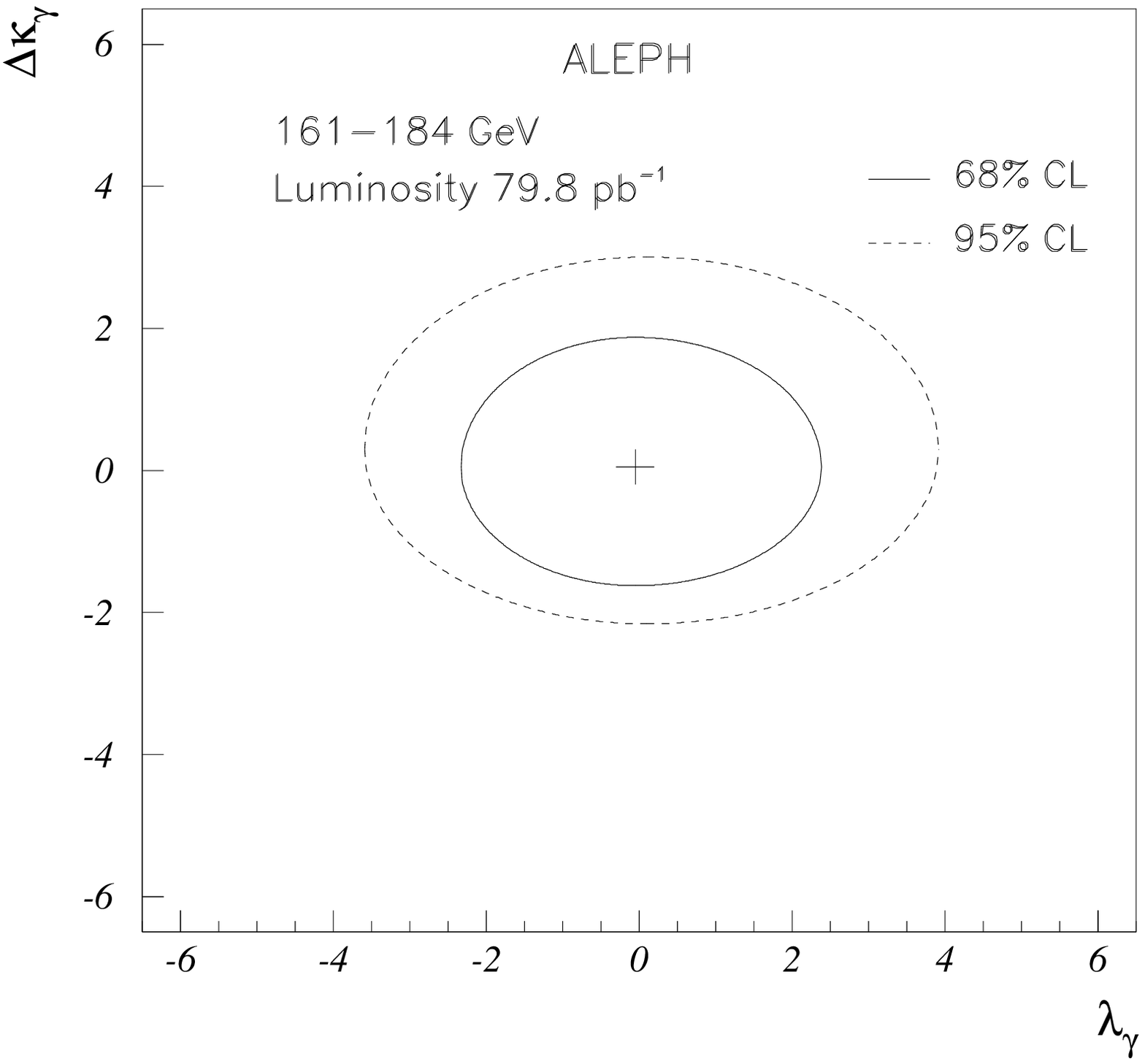,
width=11cm,height=9cm}}
\end{center}
\caption{ $68 \% $ and $95 \%$ confidence level contours in the $\deltak , \lk$ plane.}
\label{f:ellipse}
\end{figure}

  Figure \ref {f:ellipse} shows the $68 \% $ and $95 \%$ confidence level 
 contours in the ($\deltak, \lk$) plane from a two-parameter fit.
Although the two parameters are not independent, the confidence level contours are symmetric. This comes from the fact that
the results are very close to 0, so that only one minimum is found. If the results were far away from the SM prediction,
there could be several local minima, around which the two parameters would be correlated.

\section{Systematic uncertainties}

 The contributions to the systematic uncertainty on the determination of $\deltak$  are summarized in Table \ref {t:syst}.
    The total systematic uncertainty is much smaller than the statistical one.

\begin{table}[b]
\caption{Contributions to the systematic uncertainty on fitted $\deltak$, as explained in the text.}
\begin{center}
\begin{tabular}{|l|c|c|} \hline
                                      &                            &                       \\
Origin of uncertainty                 &  Region 1                  &   Region 2            \\ \hline \hline
Acceptance corrections                &  $\pm 0.08 $               &    $\pm 0.08 $        \\
Photon energy calibration $\pm1 \%$   &  $\pm 0.10 $               &    $\pm 0.20 $        \\
Background $< 1$ event                &  + 0.05                    &     + 0.05            \\
Model uncertainty $< \pm5 \%$         &  $\pm  0.10$               &   $\pm  0.15$         \\
Luminosity value          $\pm0.6 \%$ &  $\pm 0.03$                &   $\pm 0.03$          \\  \hline
Total                                 &  $\pm 0.20 $               &   $\pm 0.30 $         \\  \hline

\end{tabular}
\label{t:syst}
\end{center}
\end{table}
 
\begin{itemize}

\item  The acceptance corrections were tested with different cuts in $\xe$ and $\theta$. This led to an uncertainty on the fit
results as shown in Table 3.

\item  The main contribution to the systematic error in the present study
comes from the energy calibration of high energy photons, which 
 has been checked to be 1\% with a large sample of $\epm    \rightarrow  \gamma \gamma $ events. 

\item The possible contributions to the $\epm    \rightarrow \gamma (\gamma)  +X$ channel, other than
$X = \nu \overline{\nu}$, may come from radiative Bhabhas or 
$\epm    \rightarrow  \gamma \gamma (\gamma) $ events.
All such events in the Monte Carlo sample are eliminated
by the angular and energy cuts.

\item The KORALZ simulation of higher order effects gives a correction of about $+ 100 \%$
for the SM cross section in Region 1, and about
 $-  30 \%$  in Region 2.

A theoretical estimate of the error on these correction factors is about $5  \%$.
However, only comparisons with complete calculations from the exact matrix elements (not present in KORALZ) for the
two and three hard bremsstrahlung photons would allow a satisfactory estimation of this uncertainty.
A discussion of the uncertainty due to the implementation of the matrix elements with anomalous couplings for the
multiphoton events is presented in  \cite{reffut}.
     
The model uncertainty in introducing the anomalous couplings into the si\-mu\-lation has been checked.
The reliability of the simulation of the Standard Model is discussed in  \cite{reffut}.

\item Another contribution to the uncertainty on the total cross section part of the fit is given by the luminosity  error.
\end{itemize}

Other possible contributions to the systematic error, such as the statistical precision on the correction factors for muon
rejection and energy deposition in the forward region of the detector, are  negligible.
For $\lk$, the basic errors are the same, one region only being used for the systematic error calculation.

\section{Conclusions}
 
    The anomalous coupling parameters $\deltak$ and $\lk$ have been
measured from single and multiphoton events  in $\epm   $ collisions between 161 and 184 GeV. The results from
the fit to the cross sections and to the  energy and angular distributions of the photons are
 
\begin{eqnarray*}
       \deltak & = & \quad 0.05 ^{+1.15}_{-1.10} \mathrm{(stat)}  \pm 0.25         \mathrm{(syst)} \\
       \lk  & = & - \, 0.05  ^{+1.55}_{-1.45} \mathrm{(stat)}  \pm 0.30         \mathrm{(syst)}.
\end{eqnarray*}

The corresponding 95\% C.L. limits including systematic errors  are :

\begin{eqnarray*}
       -2.2 < \deltak < 2.3    \quad  \mathrm{assuming}  \quad \lk = 0\\
       -3.1 <  \lk <  3.2      \quad  \mathrm{assuming}  \quad \deltak=0
\end{eqnarray*}

 These results are in good agreement  with the Standard Model predictions and
    the uncertainty is largely dominated by the limited statistics of the data sample. 
 
\section{Acknowledgements}
We thank and congratulate our colleagues in the CERN accelerator divisions for the successful operation of LEP2. We are indebted
to the engineers and technicians in all our institutions for their contribution to the excellent performance of ALEPH.
Those of us from non-member countries thank CERN for its hospitality and support.
We would also like to thank J. ~Kalinowski for discussions 
and Z. W\c{a}s  for providing us with a Monte Carlo generator containing the gauge sector for the reaction
$\epm    \rightarrow \nu \overline{\nu} \gamma (\gamma)$.

\end{document}